\documentclass{PoS}
\usepackage{epsfig}

\title{Event-by-Event Fluctuations and the Search for the Critical Point within the NA49 Experiment}

\ShortTitle{Fluctuations and the Search for the Critical Point within the NA49 Experiment}

\author{\speaker{Tim Schuster} (for the NA49 Collaboration)\\
        Fachbereich Physik der Universit\"{a}t, Frankfurt, Germany.\\
        E-mail: \email{Tim.Schuster@cern.ch}}

\author{The NA49 Collaboration:\\
T.~Anticic$^{23}$, B.~Baatar$^{8}$,D.~Barna$^{4}$,
J.~Bartke$^{6}$, H.~Beck$^{9}$, L.~Betev$^{10}$, H.~Bia{\l}\-kowska$^{20}$,
C.~Blume$^{9}$,  B.~Boimska$^{20}$, J.~Book$^{9}$, M.~Botje$^{1}$,
J.~Bracinik$^{3}$, P.~Bun\v{c}i\'{c}$^{10}$,
V.~Cerny$^{3}$, P.~Christakoglou$^{1}$, P.~Chung$^{19}$, O.~Chvala$^{14}$,
J.G.~Cramer$^{16}$, P.~Csat\'{o}$^{4}$, P.~Dinkelaker$^{9}$, V.~Eckardt$^{13}$,
Z.~Fodor$^{4}$, P.~Foka$^{7}$, V.~Friese$^{7}$, J.~G\'{a}l$^{4}$,
M.~Ga\'zdzicki$^{9,11}$, V.~Genchev$^{18}$, K.~Grebieszkow$^{22}$,
S.~Hegyi$^{4}$, C.~H\"{o}hne$^{7}$, K.~Kadija$^{23}$, A.~Karev$^{13}$, D.~Kresan$^{7}$,
V.I.~Kolesnikov$^{8}$,
M.~Kowalski$^{6}$, I.~Kraus$^{7}$, M.~Kreps$^{3}$, A.~Laszlo$^{4}$, R.~Lacey$^{19}$,
M.~van~Leeuwen$^{1}$, P.~L\'{e}vai$^{4}$, L.~Litov$^{17}$, B.~Lungwitz$^{9}$,
M.~Makariev$^{18}$, A.I.~Malakhov$^{8}$, M.~Mateev$^{17}$, G.L.~Melkumov$^{8}$, C.~Meurer$^{9}$,
A.~Mischke$^{1}$, M.~Mitrovski$^{9}$, J.~Moln\'{a}r$^{4}$, St.~Mr\'owczy\'nski$^{11}$,
V.~Nicolic$^{23}$, G.~P\'{a}lla$^{4}$, A.D.~Panagiotou$^{2}$, D.~Panayotov$^{17}$,
A.~Petridis$^{2,\dagger}$, W.~Peryt$^{22}$, M.~Pikna$^{3}$, J.~Pluta$^{22}$, D.~Prindle$^{16}$,
F.~P\"{u}hlhofer$^{12}$, R.~Renfordt$^{9}$, C.~Roland$^{5}$, G.~Roland$^{5}$,
M. Rybczy\'nski$^{11}$, A.~Rybicki$^{6}$,
A.~Sandoval$^{7}$, N.~Schmitz$^{13}$, T.~Schuster$^{9}$, P.~Seyboth$^{13}$,
F.~Sikl\'{e}r$^{4}$, B.~Sitar$^{3}$, E.~Skrzypczak$^{21}$, M.~Slodkowski$^{22}$,
G.~Stefanek$^{11}$, R.~Stock$^{9}$, C.~Strabel$^{9}$, H.~Str\"{o}bele$^{9}$, T.~Susa$^{23}$,
I.~Szentp\'{e}tery$^{4}$, J.~Sziklai$^{4}$, M.~Szuba$^{22}$, P.~Szymanski$^{20}$,
M.~Utvi\'{c}$^{9}$, D.~Varga$^{4,10}$, M.~Vassiliou$^{2}$,
G.I.~Veres$^{4,5}$, G.~Vesztergombi$^{4}$, D.~Vrani\'{c}$^{7}$,
Z.~W{\l}odarczyk$^{11}$, A.~Wojtaszek$^{11}$, I.K.~Yoo$^{15}$}

\author{\\
\vspace*{1cm}\\
$^{1}$NIKHEF, Amsterdam, Netherlands. \\
$^{2}$Department of Physics, University of Athens, Athens, Greece.\\
$^{3}$Comenius University, Bratislava, Slovakia.\\
$^{4}$KFKI Research Institute for Particle and Nuclear Physics, Budapest, Hungary.\\
$^{5}$MIT, Cambridge, USA.\\
$^{6}$Henryk Niewodniczanski Institute of Nuclear Physics, Polish Academy of Sciences, Cracow, Poland.\\
$^{7}$Gesellschaft f\"{u}r Schwerionenforschung (GSI), Darmstadt, Germany.\\
$^{8}$Joint Institute for Nuclear Research, Dubna, Russia.\\
$^{9}$Fachbereich Physik der Universit\"{a}t, Frankfurt, Germany.\\
$^{10}$CERN, Geneva, Switzerland.\\
$^{11}$Institute of Physics, Jan Kochanowski University , Kielce, Poland.\\
$^{12}$Fachbereich Physik der Universit\"{a}t, Marburg, Germany.\\
$^{13}$Max-Planck-Institut f\"{u}r Physik, Munich, Germany.\\
$^{14}$Charles University, Faculty of Mathematics and Physics, Institute of Particle and Nuclear Physics, Prague, Czech Republic.\\
$^{15}$Department of Physics, Pusan National University, Pusan, Republic of Korea.\\
$^{16}$Nuclear Physics Laboratory, University of Washington, Seattle, WA, USA.\\
$^{17}$Atomic Physics Department, Sofia University St. Kliment Ohridski, Sofia, Bulgaria.\\ 
$^{18}$Institute for Nuclear Research and Nuclear Energy, Sofia, Bulgaria.\\ 
$^{19}$Department of Chemistry, Stony Brook Univ. (SUNYSB), Stony Brook, USA.\\
$^{20}$Institute for Nuclear Studies, Warsaw, Poland.\\
$^{21}$Institute for Experimental Physics, University of Warsaw, Warsaw, Poland.\\
$^{22}$Faculty of Physics, Warsaw University of Technology, Warsaw, Poland.\\
$^{23}$Rudjer Boskovic Institute, Zagreb, Croatia.\\
$^{\dagger}$deceased}

\newcommand{\snn}{\ensuremath{\sqrt{s_{\mathrm{NN}}}}}
\newcommand{\pt}{\ensuremath{p_{\mathrm{T}}}}
\newcommand{\mpt}{\ensuremath{\langle p_{\mathrm{T}} \rangle}}
\newcommand{\mub}{\ensuremath{\mu_{\mathrm{B}}}}
\newcommand{\cbs}{\ensuremath{C_{\mathrm{BS}}}}
\newcommand{\sdyn}{\ensuremath{\sigma_{\mathrm{dyn}}}}
\newcommand{\sdata}{\ensuremath{\sigma_{\mathrm{data}}}}
\newcommand{\smix}{\ensuremath{\sigma_{\mathrm{mix}}}}
\newcommand{\tch}{\ensuremath{T_{\mathrm{chem}}}}

\abstract{In heavy-ion collisions in the energy regime probed at the CERN SPS, experimental hints for the deconfinement phase transition have been seen in numerous inclusive hadronic observables. In order to further characterize this transition, and in the pursuit of indications for the expected critical point of strongly interacting matter, the NA49 collaboration has conducted analyses of the event-by-event fluctuations of various hadronic observables. A selection of these results will be presented and discussed in the light of theoretical predictions. Among these are new results on hadron ratio fluctuations, in particular K/p fluctuations and their potential connection to the correlation between strangeness and baryon number,  thus revealing the basic degrees of freedom produced in heavy-ion collisions.}

\FullConference{5th International Workshop on Critical Point and Onset of Deconfinement - CPOD 2009,\\
		 June 08 - 12 2009\\
		 Brookhaven National Laboratory, Long Island, New York, USA}

\begin{document}

\section{Introduction}

\begin{figure}
\begin{center}
\epsfig{file=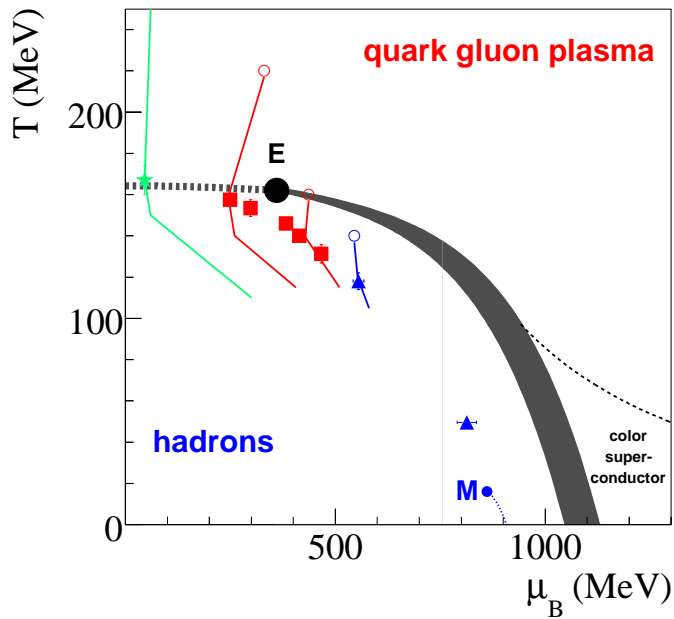,width=0.6\textwidth}
\caption{Sketch of the phase diagram of strongly interacting matter in the plane temperature ($T$) vs.\ baryonic chemical potential (\mub). Symbols denote the chemical freeze-out parameters of heavy-ion collisions at different energies as extracted by statistical model fits~\cite{Becattini:2005xt}. The line indicates the conjectured phase transition, changing from first order (full line) to a cross-over (dashed line) at the critical endpoint \emph{E}~\cite{Fodor:2004nz}.}
\label{fig_pd}
\end{center}
\end{figure}

Following the observation of a new state of matter~\cite{Heinz:2000bk} created in heavy ion collisions at the top CERN SPS energy ($\snn = 17.3~\mathrm{GeV}$), 
inclusive hadronic observables gave evidence that the onset of the deconfinement phase transition~\cite{Gazdzicki:1998vd} is observed at low SPS energies~\cite{:2007fe,Friese:CPOD09}. The energy dependence of these observables changes dramatically around $\snn = 8~\mathrm{GeV}$, most prominently the non-monotonic behavior of the average $\langle K^+ \rangle / \langle \pi^+ \rangle$ ratio~\cite{:2007fe}, and the step observed in the slope parameter of hadron transverse momentum spectra~\cite{Gazdzicki:2004ef}. In contrast, the evolution of these observables from top SPS on to RHIC energies shows no discontinuous behaviors.
Figure~\ref{fig_pd} presents a sketch of the phase diagram of strongly interacting matter, indicating the features predicted from
lattice QCD and QCD model calculations. Deconfined matter,
the QGP, is separated from the hadron phase by a first order
transition boundary at large baryo-chemical potential \mub,
ending in a critical point \emph{E} and then turning into a
cross-over transition at low values of \mub. Lattice QCD
calculations predict a critical temperature between 160 and
170 MeV at $\mub = 0$. A recent extension to the finite \mub\ domain allowed to estimate the position of the critical
point \emph{E}~\cite{Fodor:2004nz}. The locations of the hadron freeze-out points of the
high density fireball produced in nucleus-nucleus collisions
are obtained from fits of a statistical model to hadron
abundances (cf.\ e.g.~\cite{Becattini:2003wp,BraunMunzinger:2001as,Becattini:2005xt}). 
     
Here, various implementations of this model agree in the resulting chemical freeze-out conditions, and see the extracted temperature approach the cross-over temperature with rising energy. In addition, the freeze-out points for central Pb+Pb collisions in the CERN SPS energy range lie close to the critical point predicted in~\cite{Fodor:2004nz}.

The study of event-by-event fluctuations promises to convey more information about both prominent features of the phase diagram, the onset of deconfinement and the critical point.
The original assumption~\cite{Stock:1994ve} was that in heavy-ion collisions that freeze out close to the phase transition, small initial density fluctuations may lead to two distinct event classes and thus be reflected in larger event-by-event fluctuations.
In addition, fluctuations are expected to reveal information about the \emph{nature} of the phase transition and in particular about the critical point~\cite{Stephanov:1999zu,Gorenstein:2003hk}, as diverging susceptibilities near the critical point are directly connected to fluctuations.

As indicated above, the energy range covered by NA49 in the SPS energy scan ($6.3 \leq \snn \leq 17.3$~GeV) makes it possible to study an interesting range of freeze-out parameters. While quantitative predictions for the effect expected at the phase transition or the critical point are scarce, the systematic energy scan allows to search for effects that appear and/or disappear in the excitation function.
NA49's large acceptance for hadrons, as well as its independent determination of event centrality make it ideally suited for this systematic study. More experimental details can be found in~\cite{Afanasev:1999iu}.

The emphasis of this contribution is on NA49 results on the energy dependence of hadron ratio fluctuations and their interpretation as a potential signature for the onset of deconfinement and the critical point. After discussion of the $K/\pi$, $p/\pi$, and $K/p$ fluctuations, the energy and system size dependence of multiplicity $N$ and mean transverse momentum \mpt\ fluctuations are presented. These results are compared to quantitative theoretical predictions for the effect of the critical point.

For future fluctuation studies, higher moments of event-by-event distributions will be of strong interest. For this reason, this article closes with an outlook on baseline model calculations which will be of interest when measuring net baryon or net proton kurtosis.

\section{Hadron Ratio Fluctuations}

Event-by-event hadron ratios characterize the chemical composition of the fireball in each event.
Especially fluctuations of net baryon number or strangeness are sensitive to the properties of the early stage. Compared to the fluctuations of other conserved quantities (e.g.\ charge~\cite{Zaranek:2001di}), they are less strongly affected by hadronic re-interaction in later stages of the collision, so the signature of the phase transition is less prone to be washed out.
From the change in inclusive particle (e.g.\ relative strangeness) production properties observed at the phase transition, we expect distinct fluctuation patterns when the chemical freeze-out approaches the phase transition.
Besides this idea, several models suggest the study of hadron ratio fluctuations to gain further insight into the nature of the deconfinement phase transition~\cite{Csernai:1995zn,Kapusta:1986cb,Stock:1989jp}. 

These model considerations are supported by lattice QCD calculations showing a change in quark number susceptibilities~\cite{Karsch:2007dp,Karsch:2007dt} at the phase transition. Susceptibilities have a direct connection to number fluctuations: $\chi \propto \langle N^2 \rangle$. A step at the transition temperature is observed in light and strange quark number susceptibilities. While the transition is smooth at $\mub = 0$, the light quark number susceptibility diverges when approaching the critical point at higher \mub.
The changing susceptibilities could be observed in hadron number fluctuations, but ratios are more robust because they are an intensive quantity, and thus less affected by other effects like e.g.\ volume fluctuations.

No quantitative predictions for the phase transition and critical point effects on hadron ratio fluctuations exist yet, but qualitatively they must be visible when measuring hadron ratio fluctuations as a function of energy---is an effect coming and going?

An especially promising observable are the fluctuations of the $K/p$ ratio. They have been suggested as ``A Diagnostic of Strongly Interacting Matter"~\cite{Koch:2005vg} and are hoped to yield a key to the degrees of freedom in the system observed in heavy-ion collisions.

\begin{figure}
\begin{center}
\epsfig{file=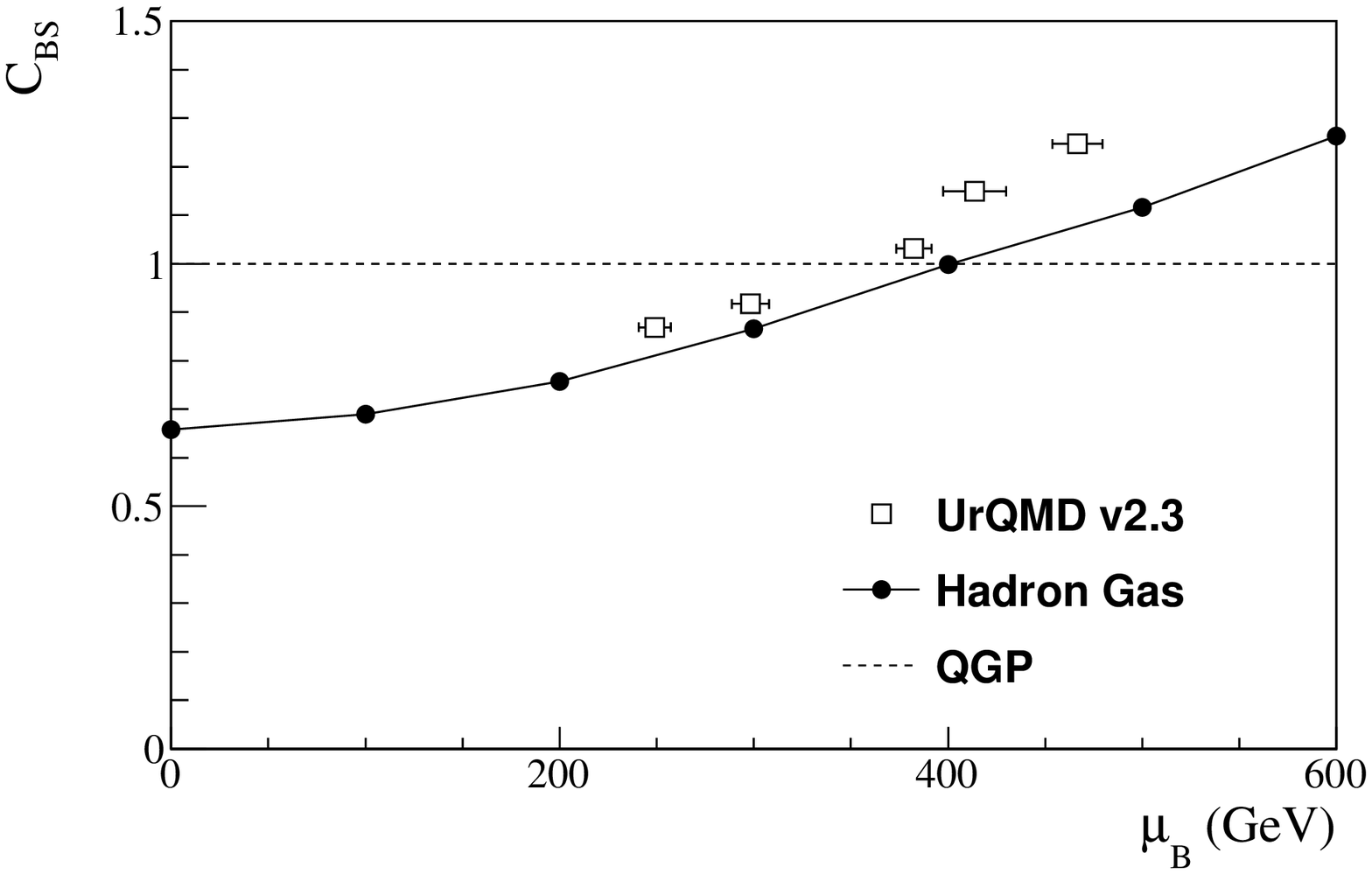,width=0.49\textwidth}
\epsfig{file=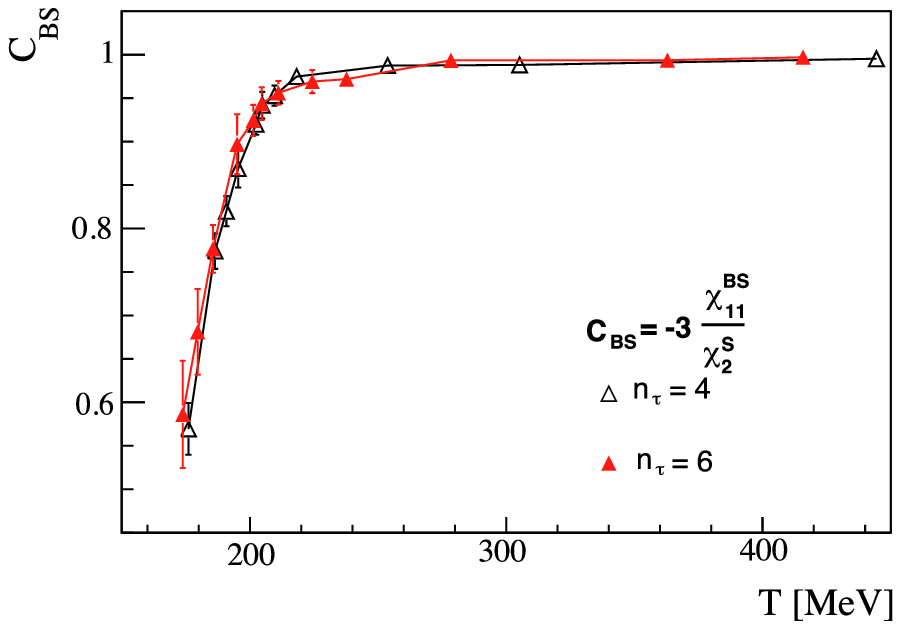,width=0.49\textwidth}
\caption{The baryon-strangeness correlation coefficient \cbs~\cite{Koch:2005vg}. Left: The expected \cbs\ for a quark gluon plasma, a grand canonical hadron-resonance gas (taken from~\cite{Koch:2005vg}) compared to calculations performed in the transport model UrQMD (see also~\cite{Haussler:2006mq}). Right: \cbs\ from a lattice QCD calculation at \mub\ = 0~\cite{Cheng:2008zh}.}
\label{fig_cbs}
\end{center}
\end{figure}

The baryon ($B$) - strangeness ($S$) correlation coefficient, defined as
\begin{equation}
\cbs = -3 \frac{\langle BS \rangle - \langle B \rangle \langle S \rangle}{\langle S^2 \rangle - \langle S \rangle^2},
\end{equation}
quantifies the correlation strength between baryon number and strangeness. This changes at the phase transition: In a quark gluon plasma, \cbs\ would be unity, as strangeness is carried by strange quarks ($S=-1$, $B=1/3$) and thus baryon number and strangeness are directly correlated.
In a hadron gas, strangeness is carried by kaons ($S=-1$, $B=0$, $\rightarrow \cbs=0$) and $\Lambda$s (S=-1, B=1, $\rightarrow \cbs=3$). A hadron gas thus shows a baryon-strangeness correlation changing with \mub. Figure~\ref{fig_cbs} (left) shows this difference. The predicted hadron gas behavior was reproduced in the hadronic transport model UrQMD. For this comparison, \cbs\ was extracted from UrQMD events at different collision energies to probe the \mub\ dependence.

As mentioned above, these observables can be directly connected to quantities measured in lattice QCD. An expression for \cbs\ in terms of susceptibilities reads~\cite{Cheng:2008zh}:
\begin{equation}
\cbs = -3 \frac{\chi^{\mathrm{BS}}_{11}}{\chi^{\mathrm{S}}_{2}}.
\label{eq_cbs_lat}
\end{equation}
Lattice QCD calculations at \mub = 0 confirm the phase transition effect~\cite{Cheng:2008zh}. Figure~\ref{fig_cbs} (right) shows the temperature dependence of \cbs\ in lattice calculations at different lattice spacings $n_{\tau}$. Recently, also the \mub\ dependence has been determined by very new, preliminary lattice calculations shown at this workshop~\cite{Schmidt:POSCPOD09}.

The definition of \cbs\ makes it necessary to measure all strange hadrons as well as all baryons event-by-event. Especially for multi-strange hyperons and neutrons this is not possible. Fluctuations in the $K/p$ ratio are thus an attempt to find a measurable proxy for the baryon-strangeness correlation. A direct, quantitative connection is, however still under discussion.

In the following, the measurement of hadron ratio fluctuations will be expressed in terms of \emph{dynamical} fluctuations~\cite{:2008ca}. The term dynamical fluctuations refers to those fluctuations remaining after
removing fluctuations from finite number statistics as well as effects from detector resolution and particle identification.
In NA49, protons, kaons and pions are identified via their energy loss in the TPC gas, and a likelihood method is used to extract the eventwise hadron ratios. A mixed event reference is subjected to the same method, so that its ratio distribution represents the finite number statistics and detector effects.
The dynamical fluctuations are defined as the quadratic difference
\begin{equation}
\sdyn = \mathrm{sign} \left(\sdata^2 - \smix^2 \right) \sqrt{ \left| \sdata^2 - \smix^2 \right| },
\label{eq_sdyn}
\end{equation}
where \sdata\ is for instance the RMS width of the event-by-event $K/\pi$ ratio normalized by the mean $\langle K/\pi \rangle$, and \smix\ is that for mixed events.
If the sign of \sdyn\ is positive, the data distribution is wider than that for mixed events
One possibility of obtaining a negative \sdyn\ (where the data distribution is narrower) is the presence of a strong correlation between the hadron species under study. Such a correlation could be a resonance that decays into the two.

A thorough description of the experimental method and the results can be found in~\cite{:2008ca}, where NA49 has recently published its final results on the energy dependence of \sdyn\ for the $K/\pi$ and $p/\pi$ ratio.
In the following, \sdyn\ for $K/\pi$, $p/\pi$ and $K/p$ is presented and compared to results at higher energies from STAR. Details on their analyses can be found in~\cite{:2009if} ($K/\pi$),~\cite{WestfallQMPoster} ($p/\pi$) and in~\cite{TianQMPoster} ($K/p$).
All data are compared to new string-hadronic transport model calculations performed in UrQMD~\cite{Bleicher:1999xi,Bass:1998ca}, in the newest version 2.3~\cite{Petersen:2008kb}.
Recent calculations of \sdyn\ ($K/\pi$) in the hadronic transport model HSD and in a statistical model~\cite{Gorenstein:2008et} are also compared to the data.

\begin{figure}
\begin{center}
\epsfig{file=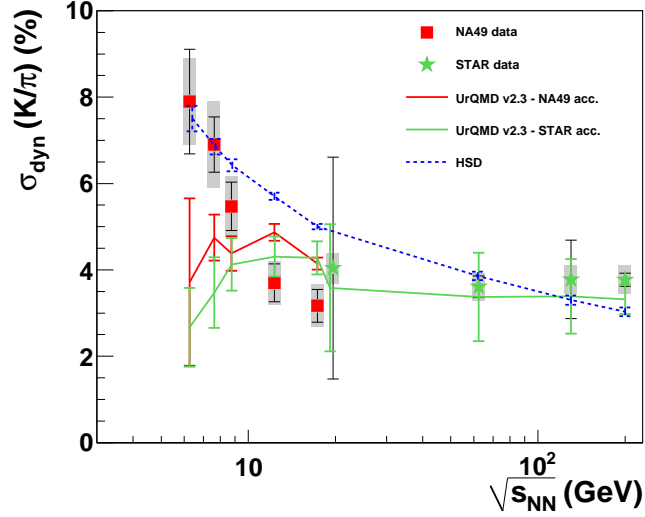,width=0.6\textwidth}
\caption{Energy dependence of \sdyn\ for the $K/\pi$ ratio. Data from NA49~\cite{:2008ca} and STAR~\cite{:2009if} is compared to transport model calculations from UrQMD and HSD.}
\label{fig_k2pi}
\end{center}
\end{figure}

Figure~\ref{fig_k2pi} shows the energy dependence of \sdyn\ for the $K/\pi$ ratio. 
The statistical errors that are indicated are mainly determined by the event statistics available at each energy. The systematic errors (indicated by shaded boxes) were deduced from systematic analysis variations~\cite{:2008ca}. 

Overall, for the $K/\pi$ ratio fluctuations, we observe positive values of \sdyn.
\sdyn\ is constant from top SPS to RHIC energies, but, towards lower energies, a strong rise is observed.
This rise is seen at the same energy where the indications for the onset of deconfinement are found in the inclusive measurements.
As no data on ratio fluctuations is available below $\snn = 6.3$~GeV, it is impossible to say whether a peak or a divergence is observed. Future programs at FAIR or NICA will answer this question.

In order to see if the observed signature is of pure hadronic origin, the data is compared to transport model calculations. For this comparison, UrQMD events have been analyzed after being subjected to acceptance filters in order to reproduce the experimental conditions. The NA49 acceptance has been applied at the SPS energies, and the full excitation function ($6.3 \leq \snn \leq 200$~GeV) has been evaluated within the STAR acceptance. These calculations are indicated in Fig.~\ref{fig_k2pi} as red and green lines, respectively. No effect from the differing acceptances is visible within UrQMD in the overlapping energy range. 

While the calculations agree with the data in the plateau region from top SPS to top RHIC energy, the rise towards low energies is not reproduced in UrQMD.
The HSD calculations~\cite{Gorenstein:2008et} have also been made in acceptances according to the experimental conditions. Their results however show a different behavior: The overall trend differs from UrQMD, although a similar idea is behind both models.
HSD describes the rise at low SPS, but fails to describe the high SPS points. Top RHIC energy results are again reproduced by the calculations.
Due to this difference, no clear conclusion can be drawn from the transport model comparison.

The authors of~\cite{Gorenstein:2008et} also used HSD to check the acceptance effect between $4\pi$ acceptance and the experimental conditions but observed no strong dependence. The $4\pi$ values were compared to statistical model calculations, which again yielded a different result and an additional dependence on the chosen statistical ensemble. Refer to~\cite{Gorenstein:2008et} and~\cite{Konchakovski:CPOD09} for a more detailed discussion of this comparison.

STAR studied the centrality dependence of $K/\pi$ fluctuations at \snn\ = 62.4 and 200 GeV~\cite{:2009if}. They use the variable $\nu_{\mathrm{dyn}}$ which is in the studied case equivalent to $\sdyn^2$ and see a scaling with the pseudo-rapidity density of produced particles at midrapidity ($\frac{\mathrm{d}N}{\mathrm{d}\eta} |_{\eta=0}$):
\begin{equation}
\nu_{\mathrm{dyn}} \approx \sdyn^2 \propto 1 / \left(\frac{\mathrm{d}N}{\mathrm{d}\eta} |_{\eta=0} \right)
\label{eq_nudyn}
\end{equation}

The NA49 data (for central collisions) does not follow the same systematics.
A different scaling law for NA49 and STAR data, taking into account the change in acceptance between the two experiments was suggested~\cite{Koch:CPOD09}. Indeed when using
\begin{equation}
\sdyn \propto \frac{1}{\sqrt{\langle K \rangle}},
\label{eq_sdscaling}
\end{equation}
where $\langle K \rangle$ is the average number of kaons within the experimental acceptance, the whole excitation function of \sdyn\ ($K/\pi$) can be reproduced.
The interpretation of this result is still debated. On the one hand, a trivial dependence on particle number would suggest that there is no unexpected, new physics behind the data. On the other hand the transport model comparison leaves an open question: While \sdyn~($K/\pi$) within HSD qualitatively shows an energy dependence like 1/$N$, such a behavior is not at all visible within UrQMD.
So although a simple explanation of the observed $K/\pi$ fluctuations has been suggested~\cite{Koch:CPOD09}, the situation still needs theoretical clarification.

At this workshop, NA49 results on the centrality dependence of \sdyn~($K/\pi$) at $\snn = 17.3$~GeV have been presented for the first time~\cite{Kresan:CPOD09}. No scaling for the energy \emph{and} the centrality dependence of \sdyn~($K/\pi$) could be found there.

\begin{figure}
\begin{center}
\epsfig{file=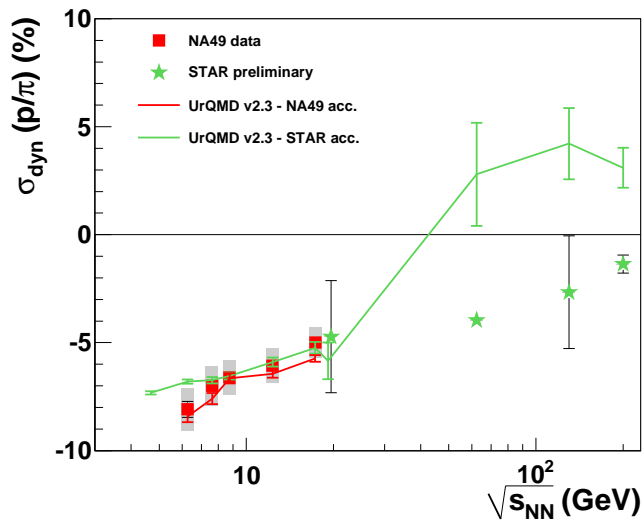,width=0.6\textwidth}
\caption{Energy dependence of \sdyn\ of the $p/\pi$ ratio. NA49 and STAR results are compared to UrQMD calculations.}
\label{fig_pr2pi}
\end{center}
\end{figure}

Looking at the available data on $p/\pi$ fluctuations (Fig.~\ref{fig_pr2pi})~\cite{:2008ca, WestfallQMPoster,Kresan:CPOD09}, we see a monotonic increase with energy from low SPS to top RHIC energies. 
The negative values are explained by a dominance of resonance decays (e.g.\ $\Delta \rightarrow p + \pi$) over fluctuations, leading to a sign opposite to that of the $K/\pi$ case~\cite{:2008ca}.

In the SPS energy range, UrQMD describes the data. When applying experimental acceptance filters on the UrQMD events, \sdyn\ does not change between NA49 and STAR acceptance. The good description through the hadronic model supports the hypothesis that the signal comes from resonance decay, thus hadronic effects.
Going to RHIC energies, a discrepancy is observed: The UrQMD results change sign and fail to describe the STAR data. It is under discussion whether this discrepancy might be due to an inadequate description of the relevant resonances in UrQMD at the high energies, or has other sources.
Another evaluation of this feature has been made at this workshop~\cite{Westfall:CPOD09}.

\begin{figure}
\begin{center}
\epsfig{file=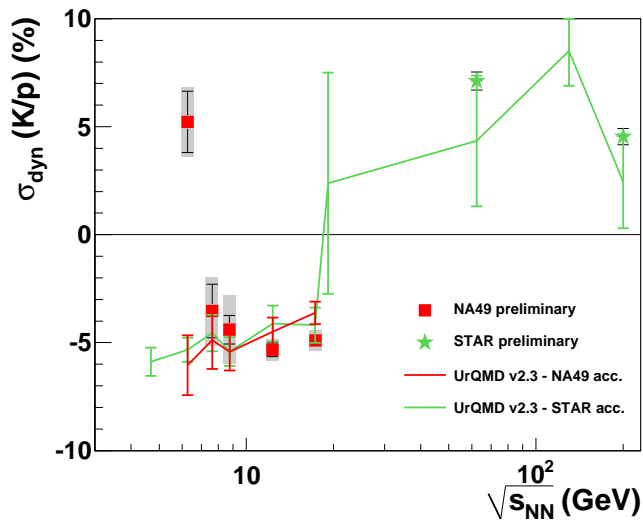,width=0.6\textwidth}
\caption{Energy dependence of \sdyn\ of the $K/p$ ratio. Data from NA49 and STAR is compared to UrQMD calculations.}
\label{fig_k2pr}
\end{center}
\end{figure}

For the first time, NA49 data on $K/p$ ratio fluctuations is presented in this contribution. Figure~\ref{fig_k2pr} shows these results together with new data from STAR~\cite{TianQMPoster}.
The most prominent feature of this measurement are the two sign changes as a function of energy. 
The negative plateau at the SPS energies from \snn\ = 7.6 to 17.3 GeV is also remarkable: no variation with energy is seen here, and negative values for \sdyn\ are observed. As this would usually be interpreted as a correlation due to resonance decay, the question arises which resonance can play a role in this case.

Between SPS and RHIC, a jump in the excitation function is followed by a weak energy dependence for the two STAR points. 
The UrQMD calculations agree well with the data from $\snn = 7.6$ to 200 GeV. The jump is also reproduced, and the absence of acceptance effects between STAR and NA49 acceptance indicates that this jump is not an effect of the changing acceptance between the two experiments.

A second jump to positive values is observed for the lowest SPS energy point. 
This first point is not explained at all by the UrQMD calculations. The UrQMD values remain negative even at $\snn = 5$~GeV. Again, further experimental measurements at even lower energies would be desirable to clarify the situation.

\section{Fluctuations at the Critical Point}

We are now looking for observables where the critical point (which is according to Fig.~\ref{fig_pd} in the neighborhood of the SPS freeze-out points) manifests itself.
At the critical point, the $\sigma$ field (the magnitude of the chiral condensate which is the order parameter of the phase transition) is expected to fluctuate wildly~\cite{Stephanov:1999zu}. 
As the $\sigma$ field is not directly measurable, we are looking for observables that convey its fluctuations to the final, detectable state: Pions couple strongly to the $\sigma$ field and we can thus expect the fluctuation pattern to be imprinted on them.
In the NA49 analysis the fluctuations in the number of charged particles is taken as a valid approximation for the number of pions.
As an observable which is directly anticorrelated to the multiplicity, the fluctuation of the mean transverse momentum \mpt\ was also studied.

NA49 has published data on the energy and system size dependence of multiplicity~\cite{Alt:2007jq} and \mpt~\cite{:2008vb} fluctuations. This data is compared to the effect of the critical point as discussed in~\cite{Stephanov:1999zu,Stephanov:pc} and at this workshop~\cite{Stephanov:CPOD09}.

When evaluating multiplicity fluctuations, it has to be kept in mind that $N$ is an extensive quantity. In order to avoid the measurements to be dominated by trivial effects, such as fluctuations in the number of participating nucleons, a strict centrality selection is applied: only the 1\% most central collisions are considered as suggested in~\cite{Konchakovski:2005hq}.
The measure for multiplicity fluctuations used in NA49 is the scaled variance (variance of the multiplicity distribution normalized by its mean):
\begin{equation}
\omega = \frac{\mathrm{Var}\left(n \right)}{\langle n \rangle} = \frac{\langle n^2 \rangle - \langle n \rangle^2}{\langle n \rangle}
\label{eq_omegadef}
\end{equation}
Following this definition, $\omega = 1$ for a Poisson distribution.

\begin{figure}
\begin{center}
\epsfig{file=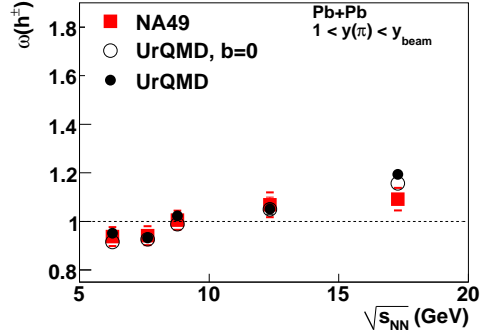,width=0.45\textwidth}
\caption{Energy dependence of multiplicity fluctuations for charged particles. The NA49 results~\cite{Alt:2007jq} are well described by the hadronic transport model UrQMD.}
\label{fig_omega_urqmd}
\end{center}
\end{figure}

The energy dependence of $\omega$ for all charged particles in the rapidity range $1 < y\left( \pi \right) < y_{\mathrm{beam}}$~\cite{Alt:2007jq} is shown in Fig.~\ref{fig_omega_urqmd}. It shows a weak increase with increasing energy and values around unity. The hadronic transport model UrQMD reproduces the observations, also for various other kinematic ranges or charge separated multiplicity fluctuations. More detailed comparisons can be found in~\cite{Alt:2007jq}.

\begin{figure}
\begin{center}
\epsfig{file=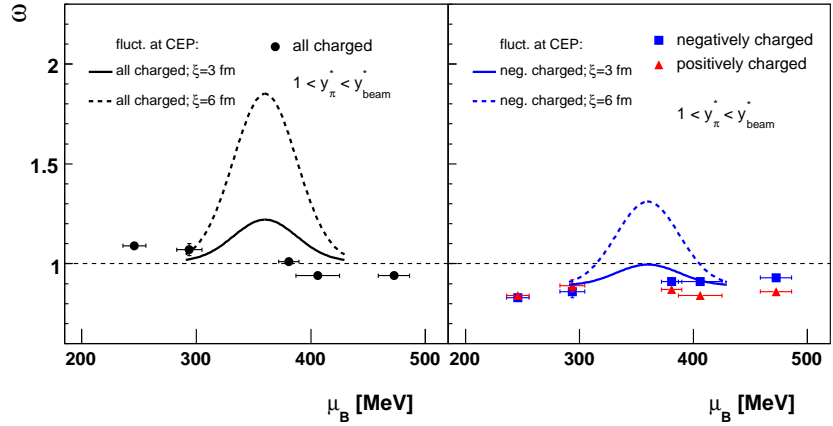,width=0.75\textwidth}
\caption{NA49 results on multiplicity fluctuations~\cite{Alt:2007jq} compared to predictions for the critical point~\cite{:2008vb,Stephanov:1999zu,Stephanov:pc}.}
\label{fig_omega_cep}
\end{center}
\end{figure}

In Figure~\ref{fig_omega_cep} (left), the same NA49 data is shown again, and in addition the values of $\omega$ for particles of only one charge are presented in the right panel. The results for different energies are plotted as a function of \mub\ where the \mub\ values were extracted from hadron gas model fits~\cite{Becattini:2005xt}.
The data is compared to the estimated effect of the critical point. The lines in Fig.~\ref{fig_omega_cep} correspond to model predictions.
While~\cite{Stephanov:1999zu,Stephanov:pc} give predictions for the amplitude of the critical point effects on multiplicity and \mpt\ fluctuations, two additional assumptions have to be made: The position of the critical point in ($T,\mub$), and the reach of the critical point effect.
For the position, an optimistic assumption is made in the following: The critical point is taken to lie directly on the freeze-out curve from~\cite{Becattini:2005xt}, while the position in \mub\ (at $\mub = 360$~MeV) is taken from lattice QCD calculations~\cite{Fodor:2004nz}. The width of the critical region is derived from the results of~\cite{Hatta:2002sj}.
The amplitude of the critical point effect is reduced by a further constraint: While in principle, the fluctuations diverge at the critical point, this is limited by the correlation length $\xi$ realized in the system. While in~\cite{Stephanov:1999zu} $\xi = 6$~fm was assumed,~\cite{Berdnikov:1999ph} expects a correlation length of $\xi = 3$~fm.
For this reason both cases were compared to the data.

The data however does not display an energy dependence that would fit to either prediction. Taking into account the very small statistical errors on the data, both cases are excluded. Another signature is also not present in the data: the amplitude of the critical point effect is expected to be twice as large for all charged particles than for separate charges.


The mean \pt\ fluctuations are quantified by the $\Phi_{\pt}$ measure~\cite{Gazdzicki:1992ri}, which represents the difference between the event average of a quantity (\mpt\ in this case) and its ensemble average, and is defined as
\begin{equation}
\Phi_{\pt} = \sqrt{\frac{\langle Z^2_{\pt} \rangle}{\langle N \rangle}} - \sqrt{\overline{\left(\pt - \overline{\pt} \right) }},
\label{eq_phidef}
\end{equation}
where
\begin{equation}
Z_{\pt} = \sum_{i=1}^{N} \left( p_{\mathrm{T}i} - \overline{\pt} \right).
\label{eq_phidef2}
\end{equation}

Uncorrelated particle production would be reflected in a value of $\Phi_{\pt} = 0$. The observable is constructed in a way to be independent of volume and multiplicity fluctuations.

\begin{figure}
\begin{center}
\epsfig{file=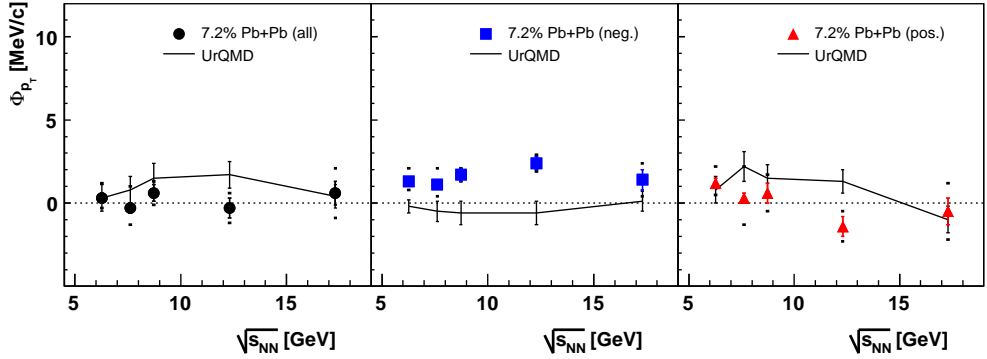,width=0.9\textwidth}
\caption{Energy dependence of \mpt\ fluctuations. The NA49 results~\cite{:2008vb} are well described by the hadronic transport model UrQMD.}
\label{fig_fipt_urqmd}
\end{center}
\end{figure}

The energy dependence~\cite{:2008vb} presented in Fig.~\ref{fig_fipt_urqmd} is flat and near zero for all charged as well as for only positive or negative particles. This behavior is, like in the case for $\omega$, reproduced by UrQMD.
An increase or peak in $\Phi_{\pt}$, as expected in the vicinity of the critical point is not visible, and no other deviation from the hadronic baseline expectations is seen.

\begin{figure}
\begin{center}
\epsfig{file=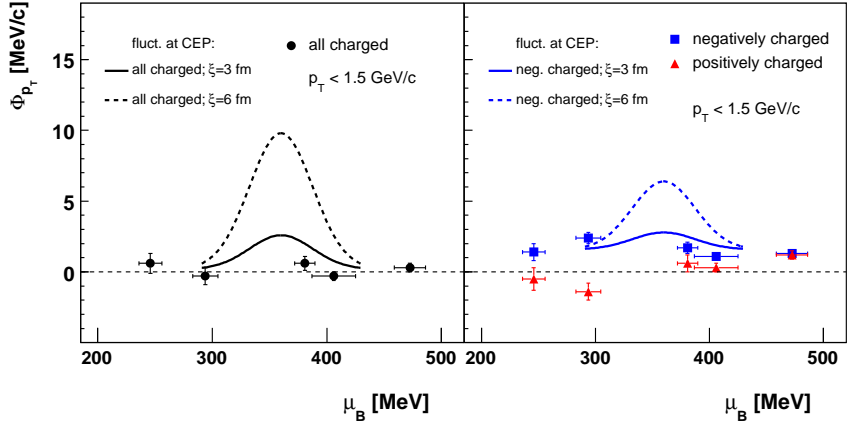,width=0.75\textwidth}
\caption{NA49 results on \mpt\ fluctuations~\cite{:2008vb} compared to expectations for fluctuations at the critical point~\cite{Stephanov:1999zu,Stephanov:pc}.}
\label{fig_fipt_cep}
\end{center}
\end{figure}

Figure~\ref{fig_fipt_cep} shows the quantitative comparison to critical point predictions, as in the case of $\omega$ plotted as a function of \mub. Again, the data exclude the predicted effects.
The absence of the effect in the data makes two conclusions possible: Either the critical point is not close enough to manifest itself in the data or the correlation length realized in heavy-ion collisions is very small.

As no significant \mub\ dependence of multiplicity and \mpt\ fluctuations was found in central Pb+Pb collisions, one is led to explore the nuclear size dependence in the search for fluctuation signatures of the critical point.
Hadron gas model fits~\cite{Becattini:2005xt} indicate that varying the collision system size at $\snn = 17.3$~GeV leads to a variation in the chemical freeze-out Temperature \tch:
for central collisions of lighter nuclei (C+C, Si+Si, p+p) the model fit obtains higher \tch\ than in central Pb+Pb collisions (the extracted freeze-out temperature for Pb+Pb is 156 MeV, while in p+p, $\tch = 180$~MeV). At the same time, almost no variation in \mub\ is observed with changing system size.
A similar pattern is seen in results of blast-wave fits to particle spectra: They yield higher \emph{kinetic} freeze-out temperatures~\cite{Gazdzicki:995681} when going to smaller systems.

These observations suggest that it might be possible to perform a 2-dimensional scan of the phase diagram of strongly interacting matter by varying \snn\ (variation of \mub) and the colliding system size $A$ (variation of $T$) in heavy-ion collisions and look for a maximum of fluctuations as a signature for the critical point.
NA49 has measured multiplicity~\cite{Alt:2006jr,benjaminPhD} and \mpt~\cite{Anticic:2003fd} fluctuations in p+p, C+C, Si+Si and Pb+Pb collisions at $\snn = 17.3$~GeV.
We compare the results to another optimistic critical point scenario: In this case the position of the critical point was chosen to lie at the same \mub\ as the 17.3 GeV freeze-out points. As the freeze-out temperatures in the smaller systems are higher the assumed $T$ of the critical point was shifted to the highest temperature freeze-out point, i.e.\ that extracted from p+p collisions: $\left( T, \mub \right) = \left( 180 \mathrm{MeV}, 250 \mathrm{MeV} \right)$.

When approaching the critical point in temperature, we now expect a rise in fluctuations. On the other hand, smaller systems must be used in order to move up in temperature. The system size begins to limit the correlation length, as the correlation length cannot exceed the system size! This results in an expectation of a maximum of fluctuations at intermediate system size.
Details of this comparison method can be found in~\cite{Grebieszkow:2009jr}.

\begin{figure}
\begin{center}
\epsfig{file=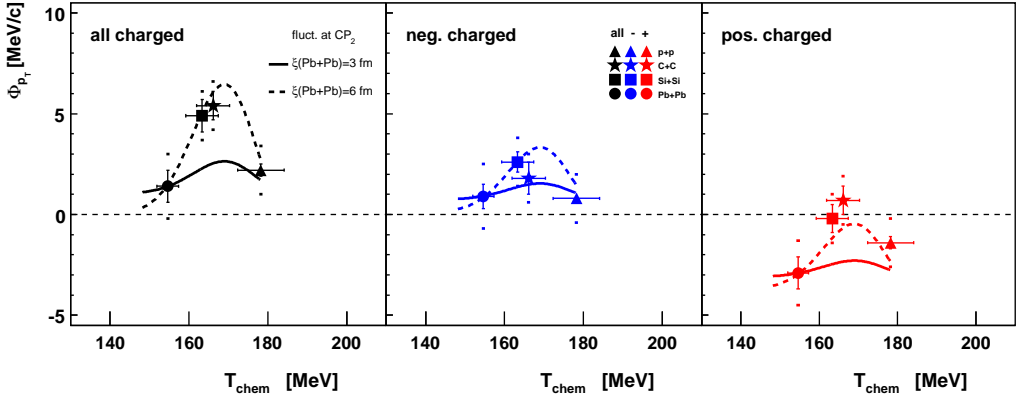,width=0.9\textwidth}
\caption{System size dependence of $\Phi_{\pt}$ at $\snn = 17.3$~GeV for p+p collisions compared to central C+C, Si+Si and Pb+Pb collisions~\cite{Anticic:2003fd}. Lines indicate expectations for the critical point~\cite{:2008vb,Grebieszkow:2009jr}.}
\label{fig_fipt_T_cep}
\end{center}
\end{figure}

We indeed see a rise in fluctuations when going to collisions of lighter ions. Figure~\ref{fig_fipt_T_cep} shows the \tch\ (system size) dependence of $\Phi_{\pt}$, where the values for Pb+Pb, Si+Si, C+C and p+p collisions are indicated at their respective chemical freeze-out temperature~\cite{Becattini:2005xt}.
The higher $\Phi_{\pt}$ results in Si+Si and C+C are compatible with the critical point scenario described above.
The prediction that the critical point effect would be twice as large in all charged particles than in only positive or negative particles is also consistent with the data.

\begin{figure}
\begin{center}
\epsfig{file=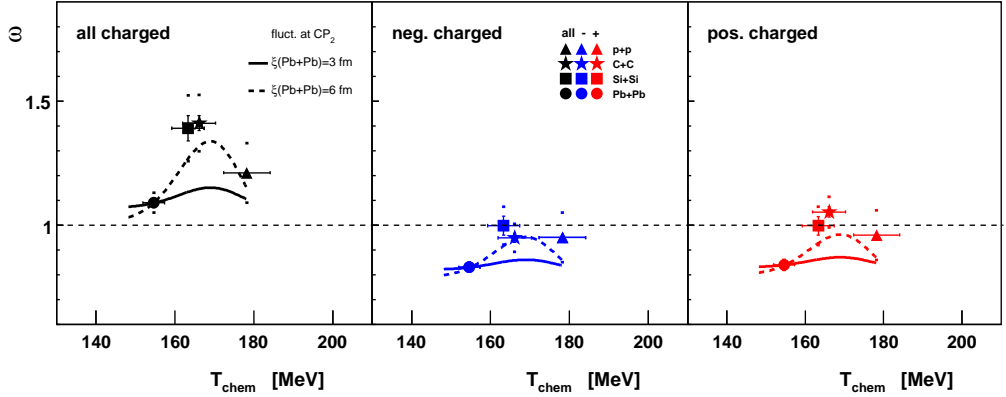,width=0.9\textwidth}
\caption{Multiplicity fluctuations ($\omega$) at $\snn = 17.3$~GeV in central p+p~\cite{Alt:2006jr}, C+C and Si+Si~\cite{benjaminPhD}, and Pb+Pb collisions~\cite{Alt:2007jq} compared to expected fluctuations from the critical point~\cite{:2008vb,Grebieszkow:2009jr}.}
\label{fig_omega_T_cep}
\end{center}
\end{figure}

Figure~\ref{fig_omega_T_cep} is the same representation for the results on $\omega$ in smaller systems. The same enhancement for intermediate system sizes is visible.

\section{Outlook: Higher Moments}
After the presentation of NA49 data, I will give as an outlook a critical assessment of a recently suggested fluctuation observable: The kurtosis of event-by-event distributions. In general, higher moments have been advertised as a sensitive probe for the phase transition~\cite{Stephanov:2008qz}. All fluctuation measures studied up to now are related to second moments.
Just like for $\omega$, a peak of fluctuations as a consequence of the diverging $\sigma$ field is also expected for the net proton skewness and kurtosis.
The advantage over second moment observables is that even in the case of a small correlation length the signal would still be sizable in the suggested higher moment observables: The amplitude of the critical point peak is proportional to higher powers of the correlation length, e.g.\ for fourth moments $\langle N^4 \rangle \propto \xi^7$, compared to $\langle N^2 \rangle \propto \xi^2$ for second moments.
Having seen above that the critical point effect may be not big enough to be visible in previously done fluctuation measurements, higher moments come up as a promising observable.
The suggested critical point effect on the kurtosis of the net proton number distribution has also been confirmed in a chiral model~\cite{Stokic:2008jh}.

In addition to the critical point effect anticipated in the net proton kurtosis, lattice QCD suggests the net baryon kurtosis as a signature of the phase transition~\cite{Cheng:2008zh}. At the transition temperature, a  step is observed.
For the experimental determination of the event-by-event net baryon distribution, neutrons are not accessible. Protons should be a valid substitute, but the expected impact of a phase transition effect on the net proton kurtosis has not been quantified yet.

In a hadronic transport model like UrQMD, it is possible to access all baryons and thus evaluate the suggested observables. This will provide a baseline on top of which, in planned future measurements, phase transition or critical point effects can be observed.
A first important finding here is that the evaluation of higher moments require larger statistics than second moments. These are of the order of millions and thus still experimentally achievable.

\begin{figure}
\begin{center}
\epsfig{file=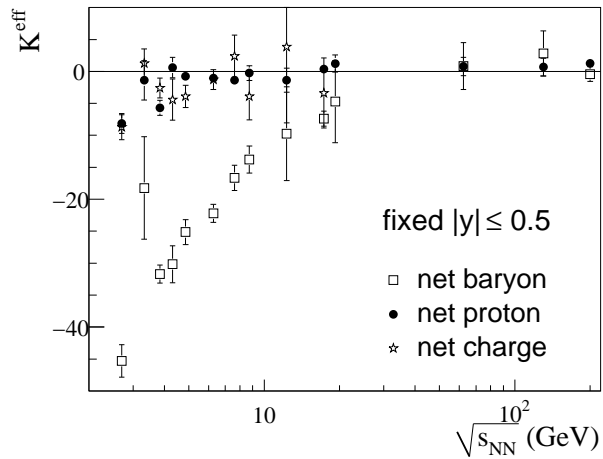,width=0.6\textwidth}
\caption{Effective kurtosis for the net-charge, net-proton and net-baryon number distributions at midrapidity ($|y|<0.5$) as calculated from UrQMD at various beam energies for central Pb+Pb/Au+Au reactions~\cite{Schuster:2009jv}.}
\label{fig_acc_kurt}
\end{center}
\end{figure}

For the UrQMD v2.3~\cite{Petersen:2008kb} study events have been analysed~\cite{Schuster:2009jv} in the energy range $2~\mathrm{GeV} < \snn  < 200~\mathrm{GeV}$. The baryon, proton and charge distributions were evaluated at a fixed acceptance around midrapidity, comparable to that of the STAR detector.
Figure~\ref{fig_acc_kurt} shows the \emph{effective kurtosis} extracted from these net baryon, net proton and net charge distributions.
The effective kurtosis of the distribution of any variable $N$ is defined as:
\begin{equation}
K^{\rm eff} = K(\delta N)\langle \delta N^2 \rangle = \frac{\langle\delta N^4\rangle}{\langle\delta N^2\rangle}-3\langle\delta N^2\rangle,
\end{equation}
where $K\left( \delta N \right)$ is the mathematical kurtosis. $K^{\rm eff}$ can directly be related to the susceptibilities that are obtained from lattice QCD~\cite{Cheng:2008zh}:
\begin{equation}
K^{\rm eff} = \frac{\chi_4}{\chi_2}.
\end{equation}

The UrQMD results show the net charge effective kurtosis fluctuating around zero. The effective kurtosis for net baryons has a strong trend towards large negative values when going to lower energies. Only at $\snn > 50$~GeV, it aproaches zero, and is thus consistent with the lattice QCD predictions which predict $K^{\rm eff} = 1$ for the hadron gas and a step to zero at the phase transition to the quark gluon plasma. The available statistics do not allow to distinguish between one and zero.
For the net protons finally the trend to negative values at low energies is also visible, but much less pronounced. At $\snn = 200$~GeV, the calculations are compatible with the first STAR results presented recently~\cite{Nayak:2009wc}.

The importance of the realistic baseline calculation is underlined in the large deviation between low temperature lattice calculations and the UrQMD results at low energies. The lattice QCD expectation of $K^{\rm eff} = 1$ for any hadron gas is in strong contrast to the negative values obtained in UrQMD. The latter can be explained by the strict quantum number conservation that is implied in UrQMD, whereas lattice QCD calculations are performed within the grand-canonical ensemble, where the expectation values for quantum numbers are controlled by chemical potentials.

While the particles within the chosen midrapidity acceptance are a valid representation of a grand-canonical ensemble at $\snn = 200$~GeV, at lower energies conservation effects become important and cause the strong deviation from lattice expectations.
This discovered background effect is much larger than the expected phase transition signature for net baryons. In net protons, this effect is much less pronounced and the expected critical point effect should not be shadowed by the background effect.

\section{Summary}
The NA49 results on the energy dependence of hadron ratio fluctuations do not give rise to a coherent interpretation: While the $p/\pi$ fluctuations can be understood in terms of resonance decay and are reproduced by hadronic models, there are contradicting interpretations of the $K/\pi$ fluctuations which remain to be settled.
The new results on $K/p$ fluctuations show a non-trivial excitation function that is not easy to understand.

A freeze-out in the vicinity of the critical point should be reflected in enhanced multiplicity or \mpt\ fluctuations. Two scenarios for critical point positions have been compared to NA49 results on $\omega$ and $\Phi_{\pt}$. 
While the results from the central Pb+Pb collision energy scan exclude a critical point located on the freeze-out curve, the enhanced fluctuations in intermediate size systems seen in NA49's system size scan at $\snn = 17.3$~GeV are consistent with the effect of a critical point located closer to the freeze-out point for collisions of lighter nuclei.
The estimated critical point effect is however strongly dependent on several model parameters.

Finally it was shown that in future fluctuation measurements using higher moments realistic baseline calculations are essential for the interpretation of the results. 

\acknowledgments
This work was supported by the US Department of Energy
Grant DE-FG03-97ER41020/A000,
the Bundesministerium f\"{u}r Bildung und Forschung, Germany (06F137), 
the Hungarian Scientific Research Foundation (T032648, T032293, T043514),
the Hungarian National Science Foundation, OTKA, (F034707),
the Polish-German Foundation, 
the Polish Ministry of Science and Higher Education (1 P03B 006 30,
1 P03B 127 30, 0297/B/H03/2007/33, N N202 078735), 
the Korea Research Foundation (KRF-2007-313-C00175) 
and the Bulgarian National Science Fund (Ph-09/05).
The author also acknowledges support by the Deutsche Forschungsgemeinschaft
(DFG) and the Helmholtz Research School on Quark Matter Studies.

\end{document}